\documentclass[a4paper,10pt]{article}
\usepackage{graphics, graphicx}
\usepackage{amssymb} 
\usepackage{latexsym}
\def\Doublehat#1{\skew3\widehat{\widehat{#1}}}
\begin{document}
\begin{center}
\LARGE  Yang-Mills Theory for Noncommutative Flows \\
\vspace{2cm}
\Large  Hiroshi TAKAI \\
\vspace{5mm}
\large  Department of Mathematics \\
Tokyo Metropolitan University \\

\end{center}
\vspace{4cm}

\begin{center}
\LARGE  Abstract 
\end{center}
\large  \quad  The moduli spaces of Yang-Mills connections on finitely
generated projective modules associated with noncommutative flows are studied. 
It is actually shown that they are homeomorphic to those on dual modules 
associated with dual noncommutative flows. Moreover the method is also 
applicable to the case of noncommutative multi-flows.

\newpage
\Large{\S1.~Introduction} \large \quad Among miscellaneous topics in super 
string theory or M-theory, one of their most important problems is concerned 
with the compactification of fields, which means that either 10 or 11 
dimensional field theory would be reduced to 4 dimensional one by compactifying either 6 or 7 dimensional space time respectively. For instance, an 11 
dimensional M-theory has a circle compactification to deduce a IIA-type super 
string theory, which describes a nonchiral field theory of closed strings due 
to BFSS ([4]). Moreover, this theory has also one more circle compactification
to deduce a IIB-type superstring theory, which describes a chiral field theory 
of closed strings via the so-called T-transformations ([5]). Recently, Connes, Douglas and Schwarz have shown that the field theory to such a 2-torus 
compactification cited above has a complete solution by taking the moduli spaces of Yang-Mills connections of appropriate modules for the gauge action of the 
2-torus on either commutative or noncommutative 2-torus ([2]). Actually, Connes and Rieffel have proved that the latter Yang-Mills moduli space is homeomorphic to the 2-torus ([3]). From this point of view, the problem of finding the 
Yang-Mills moduli space for a given smooth noncommutative dynamical system is a quite important one to determine the unified 4 dimensional field theory having 
the unique compactification. \
\quad In this paper, we present a certain duality of Yang-Mills moduli spaces for  noncommutative flows. More precisely, we show that the Yang-Mills moduli spaces for smooth noncommutative flows are homeomorphic to those for associated dual flows. This could be interpreted as no physical data is changed under dimension reduction of space time. The method itself is also applicable to noncommutative multi flows in principle. As a corollary, some basic examples are computed. \\

\Large{\S2.~Noncommutative Yang-Mills Theory} \large ~In this section, we review  the noncommutative Yang-Mills theory due to Connes-Rieffel[3].  Let $(A,G,\alpha)$ be a C$^*$-dynamical system, $A^\infty$ the set of all smooth 
elements of $A$ under $\alpha$ and $\alpha^\infty$ the restriction of $\alpha$ 
to $A^\infty$ where $G$ is a connected Lie group. Then the system 
$(A^\infty,G,\alpha^\infty)$ becomes a noncommutative smooth dynamical system. 
In what follows, we only treat such a dynamical system, so that we notationally write it by $(A,G,\alpha)$. Let $\delta$ be the differentiation map of $\alpha$. Then it is a Lie homomorphism from the Lie algebra $\mathcal{G}$ of $G$ to the Lie algebra $Der(A)$ of all $^*$-derivations of $A$. Let $\Xi$ be a finitely 
generated projective right $A$-module. Then it has a Hermitian structure 
$< \cdot \mid \cdot >_A$ with the property that
\[ < \xi \mid \eta >^*_A ~=~ < \eta \mid \xi >_A ~,\quad < \xi \mid {\eta}a >_A ~=~ < \xi \mid \eta >_A a   \]
\noindent
$(\xi,\eta \in \Xi,a \in A)$. Now we can define a noncommutative version of 
connections on vector bundles over manifolds in the following fashion: 
Let $\nabla$ be a linear map from $\Xi$ to $\Xi \otimes \mathcal{G}^*$. Then it is called a connection of $\Xi$ if it satisfies 
\[ \nabla_X({\xi}a) ~=~ \nabla_X(\xi)a ~+~ {\xi}\delta_X(a)  \]
\noindent
$(\xi \in \Xi,a \in A,X \in \mathcal{G})$. Moreover, a connection $\nabla$ is 
said to be compatible with respect to $< \cdot \mid \cdot >_A$ (or compatible) 
if it satisfies
\[ \delta_X(< \xi \mid \eta >) ~=~ < \nabla_X(\xi) \mid \eta > 
                                ~+~ < \xi \mid \nabla_X(\eta) >  \]
\noindent
$(\xi \in \Xi,a \in A,X \in \mathcal{G})$. We denote by $CC(\Xi)$ the set of all compatible connections of $\Xi$.  Then it is nonempty because it contains the 
so-called Grassmann connection $\nabla^0$, which is defined as follows: 
By assumption, $\Xi = P(A^n)$ for some $n \geq 1$ and a projection 
$P \in M_n(A)$, so that $\nabla^0 = P[\delta^n]$ becomes a compatible connection of $\Xi$, where $\delta^n$ is the differentiation map of the action $\alpha^n = \alpha \otimes id_n$ on $M_n(A)$. Now for any $\nabla \in CC(\Xi)$, there 
exists an element $\Omega_X \in E=\rm{End}_A(\Xi)$ such that
\[  \nabla_X ~=~ \nabla^0_X ~+~ \Omega_X   \]
\noindent
$(X \in \mathcal{G})$, where $\rm{End}_A(\Xi)$ is the set of all 
$A$-endomorphisms of $\Xi$. Since $\nabla$ and $\nabla^0$ are compatible, then 
$\Omega_X ~(X \in \mathcal{G})$ are all skew-adjoint. 
Given a $\nabla \in CC(\Xi)$, there exists a skew adjoint $E$-valued 2-form 
$\Theta_{\nabla}$ on $G$ such as 
\[ \Theta_{\nabla}(X,Y) ~=~ 
           \nabla_X{\nabla_Y} ~-~ \nabla_Y{\nabla_X} ~-~ \nabla_{[X,Y]}  \]
\noindent
$(X,Y \in \mathcal{G})$. It is called the curvature of $\nabla$ associated with $(A,G,\alpha)$. Then $\nabla$ is said to be flat if there exists a 2-form 
$\omega$ on $G$ such that 
\[  \Theta_{\nabla}(X,Y) ~=~ \omega(X,Y)~Id_{E}  \]
\noindent
$(X,Y \in \mathcal{G})$. We now assume an existance of a continuous $\alpha$-
invariant faithful trace $\tau$ on $A$ as anoncommutative version of
integrability of manifolds. Then there also exists a continuous faithful trace 
$\tilde{\tau}$ on $E$ such that 
\[  \tilde{\tau}(< \xi \mid \eta >_{E}) ~=~\tau(< \eta \mid \xi >_A)  \]
\noindent
$(\xi,\eta \in \Xi)$, where
\[  < \xi \mid \eta >_{E}(\zeta) ~=~ \xi < \eta \mid \zeta >_A  \]
\noindent
$(\xi,\eta,\zeta \in \Xi)$. In fact, it is well defined because of the 
assumption of $\Xi$. Using $\tilde{\tau}$, we define a noncommutative version of the Yang-Mills functional on manifolds as follows:
\[  \rm{YM}(\nabla) ~=~ - \tilde{\tau}(\{{\Theta}_{\nabla}\}^2)  \]
where
\[  \{\Theta_{\nabla}\}^2 ~=~ \sum_{i<j}~\Theta_{\nabla}(X_i \wedge X_j)^2 \in E  \]
for an orthnormal basis $\{X_i\}_i$ of $\mathcal{G}$ with respect to the Killing form. Since $\Theta_{\nabla}(X_i \wedge X_j)$ are all skew adjoint, 
then $\{\Theta_{\nabla}\}^2$ has negative spectra only. 
Therefore $\rm{YM}(\nabla) \geq 0$ for all $\nabla \in CC(\Xi)$. Moreover,
it is independent of the choice of a hermitian structure $<\cdot \mid \cdot>_A$
on $\Xi$. Now let $U(E)$ be the set of all unitaries of $E$. 
It is called the gauge group of $\Xi$. For any $u \in U(E)$, we define the gauge transformation $\gamma_u$ on $CC(\Xi)$ by
\[  (\gamma_u(\nabla))_X(\xi) ~=~ u(\nabla_X)(u^*\xi)   \]
\noindent
$(u \in U(E),X \in \mathcal{G},\xi \in$ $\Xi$). Then $\gamma$ calls the gauge action of $U(E)$ on $CC(\Xi)$. The Yang-Mills functional $\mathrm{YM}$ is $\gamma$-invariant, namely
\[  \mathrm{YM}(\gamma_u(\nabla)) = \mathrm{YM}(\nabla)  \]
\noindent
$(u \in U(E),\nabla \in CC(\Xi))$. We then consider the first variational problem of $\mathrm{YM}$, namely find a $\nabla \in CC(\Xi)$ such that 
\[  \frac{d}{dt}(\mathrm{YM}(\nabla_t))\Bigm{\vert}_{t=0} ~=~ 0   \]
for any smooth path $\nabla_t \in CC(\Xi)~(\vert t \vert<\epsilon)$ with 
$\nabla_0=\nabla$, which is called a Yang-Mills connection of $\Xi$ with 
respect to the system $(A,G,\alpha,\tau)$. Let $MC(\Xi)$ be the set of all Yang-Mills connections of $\Xi$ with respect to $(A,G,\alpha,\tau)$. Then the orbit 
space $\mathcal{M}^{(A,G,\alpha,\tau)}(\Xi)$ of $MC(\Xi)$ by the gauge action 
$\gamma$ of $U(E)$ is called the moduli space of the Yang-Mills connections of 
$\Xi$ with respect to the system $(A,G,\alpha,\tau)$. 
We then state the following theorem due to Connes-Rieffel[3] which is quite 
powerful to construct a Yang-Mills connection: \\

Theorem 2.1([3]) \quad Let $(A,G,\alpha)$ be a $C^{\infty}$-dynamical system and $\tau$ be a faithful $\alpha$-invariant continuous trace on $A$. Let $\Xi$ be a finitely generated projective right $A$-module. If G is an abelian connected 
Lie group, then $\nabla \in MC(\Xi)$ if and only if it is flat for any $\nabla 
\in CC(\Xi)$. \\

\Large{\S3.~Dual Yang-Mills Moduli spaces } \large \quad 
In this section, we only take Frechet flows (or multi-flows) as a special case 
of C$^{\infty}$-dynamical systems. According to Elliott-Natsume-Nest[7], let $(A,\mathbb{R},\alpha)$ be a Frechet $^*$-flow in the sence that 
\begin{enumerate}
  \item[(1)]  $(A,\{\| \cdot \|_n\}_{n \geq 1})$ is a Frechet $^*$-algebra
  			  (which is dense in a C*-algebra),
  \item[(2)]  $t \mapsto \alpha_t(a)$ is $C^{\infty}$-class with respect to
              $\| \cdot \|_n$ ($n \geq 1$), 
  \item[(3)]  For any $m,k\geq 1$, there exist $n,j\geq 1$ and $C>0$ such that
\[ \Bigm \| \frac{d^k}{dt^k} \alpha_t(a) \Bigm \|_m  \leq 
      C(1+t^2)^{j/2}  \|a\|_n \qquad (a \in A,~t \in \mathbb{R})  \]
\end{enumerate}
In what follows, we state Frechet $^*$-flows by F$^*$-flows. Typical are the 
following three examples as F$^*$-flows: \\

Examples 3.1 \quad Let $\mathcal{S}(\mathbb{R})$ be the abelian F$^*$-algabra of all complex valued rapidly decreasing smooth functions on $\mathbb{R}$ and 
$\lambda$ the shift action of $\mathbb{R}$ on $\mathcal{S}(\mathbb{R})$. Then 
the triplet $(\cal{S}(\mathbb{R}),\mathbb{R},\lambda)$ is a F$^*$-flow. \\

Example 3.2 \quad Let $\mathcal{K}^\infty(\mathbb{R})$ be the F$^*$-algebra 
consisting of all compact operators on $L^2(\mathbb{R})$ with their integral 
kernels in $\mathcal{S}(\mathbb{R}^2)$, and $Ad(\lambda)$ the adjoint action of $\mathbb{R}$ on $\mathcal{K}^\infty(\mathbb{R})$. Then the triplet $(\mathcal{K}^\infty(\mathbb{R}),\mathbb{R}$,$Ad(\lambda)$) is a F$^*$-flow. \\

Example 3.3 \quad Let $\mathbb{R}_{\theta}^2$ be the F$^*$-algebra 
$\mathcal{S}$($\mathbb{R}^2)$ with Moyal product $\star_{\theta}~(\theta \in \mathbb{R})$, and $\widehat{\theta}$ the dual action of the canonical action $\theta$ on $\mathcal{S}(\mathbb{R})$. Then the triplet $(\mathbb{R}_{\theta}^2,\mathbb{R},\widehat{\theta})$ is a F$^*$-flow. \\

Now let $(A,\mathbb{R},\alpha)$ be a F$^*$-flow with a continuous $\alpha$-
invariant faithful trace $\tau$, and let $\mathcal{S}(\mathbb{R},A)$ be the 
F$^*$-algebra consisting of all $A$-valued rapidly decreasing smooth functions 
on $\mathbb{R}$ with its seminorms $\|\cdot\|_{m,n}$ given by
\[  \| x \|_{m,n} = \mathrm{sup}_{t \in \mathbb{R}}(1+t^2)^{m/2} 
		\Bigm \| \frac{d^n}{dt^n} x(t) \Bigm \|_m    \]
\noindent
($x \in \mathcal{S}(\mathbb{R},A)$). Moreover it has the following product and 
involution:
 \[   (1)~ (x *_\alpha y)(t) = 
		 \int_\mathbb{R} x(s)\alpha_s(y(t-s))ds ~, ~
	        (2)~ x^*(t) = \alpha_t(x(-t))^*   \]
\noindent	    
($x,y \in \mathcal{S}(\mathbb{R},A)$). Then we call $\mathcal{S}(\mathbb{R},A)$ the F$^*$-crossed product of $A$ by the action $\alpha$ of $\mathbb{R}$, which 
is written by $A\rtimes_\alpha \mathbb{R}$. In fact, the three examples cited 
above are isomorphic to $\mathbb{C}\rtimes_\iota \mathbb{R}$, 
$\mathcal{S}(\mathbb{R})\rtimes_\lambda \mathbb{R}$ and 
$\mathcal{S}(\mathbb{R})\rtimes_\theta \mathbb{R}$ respectively. Then we define two actions $\widehat{\alpha}, \widetilde{\alpha}$ of $\mathbb{R}$ on 
$A \rtimes_\alpha \mathbb{R}$ given by 
\[\widehat{\alpha}_s(x)(t)=e^{2{\pi}ist}x(t),~
  \widetilde{\alpha}_s(x)(t)=\alpha_s(x(t)) \quad (i=\sqrt{-1})  \]
\noindent
$(x \in A\rtimes_\alpha \mathbb{R},~s,t \in \mathbb{R})$. The triplets $(A\rtimes_\alpha \mathbb{R},\mathbb{R},\widehat{\alpha})$ and $(A\rtimes_\alpha \mathbb{R},\mathbb{R},\widetilde{\alpha})$ become F$^*$-flows. The former is called to 
be the dual F$^*$-flow of $(A,\mathbb{R},\alpha)$. Then the same duality holds as in the case of $C^*$-crossed products in the following: \\

Theorem 3.4([7]) \quad Given a F$^*$-flow $(A,\mathbb{R},\alpha)$, its double 
dual F$^*$-flow $(A \rtimes_\alpha \mathbb{R} \rtimes_{\widehat{\alpha}} 
\mathbb{R},\mathbb{R},\Doublehat{\alpha})$ is isomorphic to the F$^*$-flow 
$(A \otimes \mathcal{K}^\infty(\mathbb{R}),\mathbb{R},\alpha \otimes Ad(\lambda))$. \\
\noindent
In fact, the equivariant isomorphism $\Psi_{\alpha}^0: 
A \rtimes_\alpha \mathbb{R} \rtimes_{\widehat{\alpha}} \mathbb{R}
\longmapsto A \otimes \mathcal{K}^\infty(\mathbb{R})$ is given by
\[ \Psi_{\alpha}^0(x)(t,s) = 
			\int_{\mathbb{R}} e^{2{\pi}irs}\alpha_{-t}(x(t-s,r))~dr  \]
\noindent
$(x \in A \rtimes_{\alpha} \mathbb{R} \rtimes_{\widehat{\alpha}}\mathbb{R},~t,s  \in \mathbb{R})$. Then the inverse isomorphism $(\Psi_{\alpha}^0)^{-1}$ of 
$\Psi_{\alpha}^0$ is given by
\[ (\Psi_{\alpha}^0)^{-1}(x)(t,s) = 
		\int_{\mathbb{R}} e^{2{\pi}i(t-r)s}\alpha_r(x(r,r-t))~dr   \]
\noindent
$(x \in A \otimes \mathcal{K}^\infty(\mathbb{R}),~t,s \in \mathbb{R})$. 
Now let $(A \rtimes_\alpha \mathbb{R},\mathbb{R},\widehat{\alpha})$ be the dual F$^*$-flow of $(A,\mathbb{R},\alpha)$. If there exists a continuous faithful 
$\alpha$-invariant trace $\tau$ on $A$, then so does it for the F$^*$-flow 
$(\widehat{A},\mathbb{R},\widehat{\alpha})$ given by
\[ 	 \widehat{\tau}(x) ~=~ \tau(x(0))   \qquad (x \in \widehat{A})  \]
\noindent
where $\widehat{A} = A \rtimes_\alpha \mathbb{R}$. Then $\widehat{\tau}$ is 
called the dual trace of $\tau$. Then we consider the Yang-Mills moduli spaces 
for such dual systems. Namely, let $\Xi$ be a finitely generated projective 
right $A$-module and $\widehat{\Xi} = \mathcal{S}(\mathbb{R},\Xi)$ the set of 
all $\Xi$-valued rapidly decreasing smooth functions on $\mathbb{R}$. Then it 
becomes a finitely generated projective right $\widehat{A}$-module. Indeed, the action of $\widehat{A}$ on $\widehat{\Xi}$ is given by 
\[	({\xi}x)(t)=\int_\mathbb{R} \xi(s)\alpha_s(x(t-s))~ds	\]
\noindent
$(\xi \in \widehat{\Xi},x \in \widehat{A})$. On the other hand, the action 
$\widetilde{\alpha}$ is implimented by an unitary multiplier flow on 
$\widehat{A}$, namely there exists a strictly continuous unitary flow $\widetilde{u}$ of the multiplier algebra of $\widehat{A}$ such that $\widetilde{\alpha}_t=Ad(\widetilde{u}_t)$ on $\widehat{A}$. Then it follows that $\widehat{\tau}$ is $\widetilde{\alpha}$-invariant. Since the action $\widehat{\alpha}$ commutes 
with $\widetilde{\alpha}$, we can define the action $\overline{\alpha}$ on $\widehat{A}$ by $\widehat{\alpha} \circ \widetilde{\alpha}$ which makes $\widehat{\tau}$ invariant. Then we propose anothet Yang-Mills moduli space 
$\mathcal{M}^{(\widehat{A},\mathbb{R},\overline{\alpha},\widehat{\tau})}(\widehat{\Xi})$ of $\widehat{\Xi}$ with respect to the F$^*$-flow $(\widehat{A},\mathbb{R},\overline{\alpha})$ and the dual trace $\widehat{\tau}$, 
which is called the dual Yang-Mills moduli space of $\mathcal{M}^{(A,\mathbb{R},\alpha,\tau)}(\Xi)$. \\

\Large{\S4.~Main result} \large \quad 
In this section, we prove the following theorem, 
which means physically that in quantum field theory, physical data 
are invariant under dimension reduction: \\

Theorem 4.1 \quad Let $(A,\mathbb{R},\alpha)$ be a F$^*$-flow 
with a continuous $\alpha$-invariant faithful trace $\tau$ and let $\Xi$ be a 
finitely generated projective right $A$-module. 
Then there exist a F$^*$-flow 
$(\widehat{A},\mathbb{R},\overline{\alpha})$ with a dual trace 
$\widehat{\tau}$ of $\tau$ and a finitely generated projective right 
$\widehat{A}$-module $\widehat{\Xi}$ whose Yang-Mills moduli space 
${\mathcal{M}}^{(\widehat{A},\mathbb{R},\overline{\alpha},\widehat{\tau})}
(\widehat{\Xi})$ is homeomorphic to 
$\mathcal{M}^{(A,\mathbb{R},\alpha,\tau)}(\Xi)$~. \\

\noindent
Applying Theorems 3.4 and 4.1, we have the following corollary: \\

Corollary 4.2 \quad  Let $(\widehat{A},\mathbb{R},\overline{\alpha})$ be the F$^*$-flow cited in Theorem 4.1 and $\beta$ a smooth action commuting with $\overline{\alpha}$. Suppose there exists a continuous faithful $\beta$-invariant trace $\tau$, then given a finitely generated projective right $\beta$-module $\Xi$, 
there exists a F$^*$-flow $(A,\mathbb{R},\beta_A)$ with a continuous faithful 
$\beta_A$-invariant trace $\tau_A$ and a finitely generated projective 
$A$-module $\Xi_A$ such that 
$\mathcal{M}^{(\widehat{A},\mathbb{R},\beta,\tau)}(\Xi)$ is homeomorphic to 
$\mathcal{M}^{(A,\mathbb{R},\beta_A,\tau_A)}(\Xi_A)~.$ \\

Proof of Theorem 4.1: \quad By the assumption of $\Xi$, there exist a natural 
number $n$ and a projection $P \in M_n(A)$ such that $\Xi = P(A^n)$. 
Let us take a Hermitian structure $<\cdot \mid \cdot >_A$ on $\Xi$ by
\[ < \xi \mid \eta >_A  ~=~ 
			\sum^n_{j=1} ~ {\xi}_j^* ~{\eta}_j    \]
\noindent
$(\xi,\eta \in \Xi)$. 
Then if $\nabla^0$ is the Grassmann connection of $\Xi$, then it belongs to 
$CC(\Xi)$. Moreover it follows from Theorem 2.1 that $\nabla^0 \in MC(\Xi)$. 
Now for any $\nabla \in MC(\Xi)$ and $X \in Lie(\mathbb{R})$, there exists a 
skew adjoint element $\Omega_X \in E$ such that 
\[ \nabla_X = \nabla^0_X  + \Omega_X ~. \]  
As $\Xi = P(A^n)$, it follows that 
$\widehat{\Xi} = \widehat{P}(\widehat{A}^n)$ 
where $\widehat{P} = P \otimes I_{\cal{S}(\mathbb{R})}$. 
Then we know that
\[\rm{End}_{\widehat{A}}(\widehat{\Xi})=\widehat{P}M_n(\widehat{A})\widehat{P}~, \]
which is denoted by $\widehat{E}$. From now on, we want to define a mapping from $\mathcal{M}^{(A,\mathbb{R},\alpha,\tau)}(\Xi)$ into $\mathcal{M}^{(\widehat{A},\mathbb{R},\overline{\alpha},\widehat{\tau})}(\widehat{\Xi})$ in the following way: \quad Since $E$ is no longer $\alpha^n$-invariant in general, it follows 
using the same idea in Connes[1] that there exists a F$^*$-flow 
$(M_n(A),\mathbb{R},\beta)$ with the property that 
\begin{enumerate}
\item [(1)] \quad $\beta_t(P)=P ~(t \in \mathbb{R})$,
\item [(2)] \quad $(M_n(A),\mathbb{R},\beta)$ is outer equivalent to 
$(M_n(A),\mathbb{R},\alpha^n)$. 
\end{enumerate} 
By [1], let $\iota_u$ be the equivariant isomorphism from the F$^*$-system 
$(M_n(A) \rtimes_{\alpha^n} \mathbb{R},\widehat{\alpha^n})$ onto $(M_n(A) 
\rtimes_{\beta} \mathbb{R}, \widehat{\beta})$ such that
\[\iota_u \circ \widetilde{\alpha^n} \circ \iota_u^{-1}=\widetilde{\beta}~.\]
\noindent
Then we have the following lemma which would be applied later: \\

Lemma 4.3([1])  \quad The next two statements holds:
\begin{enumerate}
  \item[(1)] There is an equivariant isomorphism $\widetilde{\iota_u}$ from \\
   $(M_n(A) \rtimes_{\alpha^n} \rtimes_{\widehat{\alpha^n}} \mathbb{R},\mathbb{R},\Doublehat{\alpha^n})$ onto 
$(M_n(A) \rtimes_{\beta} \mathbb{R} \rtimes_{\widehat{\beta}},\mathbb{R},\Doublehat{\beta})$.
  \item[(2)]  There exists a unitary multiplier $W$ of 
$M_n(A) \otimes \mathcal{K}^\infty(\mathbb{R})$ such that 
\[ Ad(W) \circ \Psi_{\beta}^0 \circ \widetilde{\iota_u} = \Psi_{\alpha^n}^0 ,\]
\end{enumerate} 				
where $\Psi_{\{ \cdot \}}^0$ are the equivariant isomorphisms as in Theorem 3.4 associated with $\{ \cdot \}$,and $\widetilde{\iota_u}(a)(s)=\iota_u\{a(s)\}$ 
for all $a \in \mathcal{S}(\mathbb{R},M_n(A) \rtimes_{\alpha^n} \mathbb{R})$~.\\
\noindent
Since $M_n(\Xi)=(P \otimes I_n)([M_n(A)]^n)$, it is a finitely generated 
projective $M_n(A)$-module. Let $d\beta$ be the infinitesimal generator of 
$\beta$. Now for any $\nabla \in \mathrm{MC}^{(M_n(A),\mathbb{R},\beta,\tau^n)}(M_n(\Xi))$, there exists a skew adjoint element $\Omega \in E_n = \mathrm{End}_{M_n(A)}(M_n(\Xi))$ such that
\[ \nabla = (P \otimes I_n)d{\beta}^n + \Omega . \]
Let $\widehat{E_n}=\widehat{\mathrm{End}}_{M_n(A)}(M_n(\Xi))$ be the set of all $E_n$-valued rapidly decreasing smooth functions on $\mathbb{R}$. Then it 
becomes a F$^*$-algebra with respect to the $\beta^n$-twisted convolution 
product. By definition, we see that 
\[ \widehat{E_n}=\mathrm{End}_{\widehat{M_n(A)}}(\widehat{M_n(\Xi)}) , \]
\noindent
where $\widehat{M_n(A)}=M_n(A) \rtimes_{\beta} \mathbb{R}$ and 
$\widehat{M_n(\Xi)}=\mathcal{S}(\mathbb{R},M_n(\Xi))$. Let us define the element ${\widehat{\Omega}} \in \widehat{E_n}$ by
\[ {\widehat{\Omega}}(\xi)(t)=\Omega\{\xi(t)\} \]
\noindent
$(\xi \in \widehat{M_n(\Xi)},t \in \mathbb{R})$  In fact, we check that
\[{\widehat{\Omega}}({\xi}a)(t)= \Omega\{({\xi}a)(t)\}
=\int_{\mathbb{R}}~\Omega\{(\xi)(s)\beta_s(a(t-s))\}~ds  \]
\[~~~~~~~~~~~~~~~~~~~~~~~~~~~
=\int_{\mathbb{R}}~\Omega\{\xi(s)\}\beta_s(a(t-s))~ds \]
\[~~~~~~~~~~~~~~~~~~~~~~~~~~~
=\int_{\mathbb{R}}~\widehat{\Omega}(\xi)(s)\beta_s(a(t-s))~ds  \]
\noindent
$~~~~~~~~~~~~~~~~~~~~~~~~~~~~~~~~~~~~~=({\widehat{\Omega}}({\xi}a)(t)$  \\
\noindent
$(\xi \in \widehat{M_n(\Xi)},a \in \widehat{M_n(A)})$. 
As $\beta^n$ is used as the restriction of the natural extension of $\beta$ of 
$M_{n^2}(A))$ to $E_n$, then it follows from the definition that 
\[ \widehat{E_n}=E_n \rtimes_{\beta^n} \mathbb{R} .\] 
Then we obtain that 
\[{\widehat{\Omega}} \in E_n \rtimes_{\beta^n} \mathbb{R}. \]
\noindent
We then have the following lemma:\\

Lemma 4.4 \qquad $\widehat{\Omega} \in \widehat{E_n}$ is skew adjoint.\\

Proof. \quad Since $\Omega \in E_n$ is skew adjoint, we compute that\\

$<\widehat{\Omega}(\xi \otimes f) \mid \eta \otimes g>_{\widehat{M_n(A)}}$
\[=\displaystyle \sum_{j=1}^{n} \widehat{\Omega}(\xi \otimes f)_j^*(\eta \otimes g)_j \].
\noindent
where $\widehat{\Omega}(\xi \otimes f)_j,(\eta \otimes g)_j \in \widehat{M_n(A)}$. Then it is easy to check that
\[ \widehat{\Omega}(\xi \otimes f)_j=\Omega(\xi)_j \otimes f \],
\noindent
using which we deduce that\\

$<\widehat{\Omega}(\xi \otimes f) \mid \eta \otimes g>_{\widehat{M_n(A)}}(t)$
\[=\displaystyle \sum_{j=1}^{n}{\int}_{\mathbb{R}}\beta_s(\Omega(\xi)_j^*\eta_j)\overline{f(-s)}g(t-s)~ds \]
\[~~~~~~~~={\int}_{\mathbb{R}}\beta_s\{<\Omega(\xi) \mid \eta>_{M_n(A)}\}\overline{f(-s)}g(t-s)~ds \]
\noindent
As $\Omega$ is skew adjoint, it follows that
\[<\Omega(\xi) \mid \eta>_{M_n(A)}=-<\xi \mid \Omega(\eta)>_{M_n(A)}\]. 
\noindent
Then we obtain that
\[<\widehat{\Omega}(\xi \otimes f) \mid \eta \otimes g>_{\widehat{M_n(A)}}(t)
=-<\xi \otimes f \mid \widehat{\Omega}(\eta \otimes g)>_{\widehat{M_n(A)}}(t)\]
\noindent
$(\xi,\eta \in M_n(\Xi),~f,g \in \cal{S}(\mathbb{R}))$. This implies the 
conclusion.  \quad Q.E.D. \\

Let $d\widehat{\beta^n}$ and $d\widetilde{\beta^n}$ be the infinitesimal generators of the dual action $\widehat{\beta^n}$ and the canonical extension 
$\widetilde{\beta^n}$ of $\beta^n$ to $\widehat{E_n}$ respectively. Since 
$\widehat{\beta^n}$ commutes with $\widetilde{\beta^n}$, then $d\widehat{\beta^n}+d\widetilde{\beta^n}$ is the infinitesimal generator of $\overline{\beta^n}$. Then we have the following lemma:\\

Lemma 4.5 
\[ (\widehat{P} \otimes I_n)(d\overline{\beta^n})+\widehat{\Omega} \in 
\mathrm{MC}^{(\widehat{M_n(A)},\mathbb{R},\overline{\beta},\widehat{\tau^n})}
(\widehat{M_n(\Xi)}) ,  \]

Proof. \quad Since $\widehat{P} \otimes I_n)(d\overline{\beta^n})$ is the 
Grassmann connection of $\widehat{M_n(\Xi)}$ with respect to the action 
$\overline{\beta}$, it belongs to ${\mathrm{MC}}^{(\widehat{M_n(A)},\mathbb{R},\overline{\beta},\widehat{\tau^n})}(\widehat{M_n(\Xi)})$ by Theorem 2.1. As 
$\widehat{\Omega} \in \widehat{E_n}$ is skew adjoint by Lemma 4.4, the 
conclusion follows from Theorem 2.1.  \quad Q.E.D.  \\

\noindent 
By the Lemma 4.5, we then define a mapping 
\[\Phi_{\beta}:\mathcal{M}^{(M_n(A),\mathbb{R},\beta,\tau^n)}(M_n(\Xi)) 
\longmapsto \mathcal{M}^{(\widehat{M_n(A)},\mathbb{R},\overline{\beta},
\widehat{\tau^n})}(\widehat{M_n(\Xi)})  \]
\noindent
by the following fashionF
\[ \Phi_{\beta}([\nabla]_{U(E_n)}) 
= [(\widehat{P} \otimes I_n)(d\overline{{\beta}^n})+\widehat{\Omega}~]
_{U(\widehat{E_n})},\]
\noindent
where $[\nabla]_{\{*\}}$ means the equivalence class of $\nabla$ under the gauge action of $\{*\}$. \\

We then check the following lemma: \\

Lemma 4.6 \qquad  $\Phi_{\beta}$ is well defined. \\

Proof. \quad Let $\Omega,~\Omega_{1}$ be two skew adjoint elements in $E_n$, 
and suppose $u(\nabla^0_{\beta}+\Omega)u^*=\nabla^0_{\beta}+\Omega_{1}$ 
for some unitary $u \in E_n$, then 
\[\Omega_{1}=u\nabla^0_{\beta}u^*-\nabla^0_{\beta}+u{\Omega}u^* \]
\noindent
We have to show that
$\nabla^0_{\overline{\beta}}+\widehat{\Omega}$ 
is equal to 
$\nabla^0_{\overline{\beta}}+\widehat{\Omega}_{1}$ 
up to the gauge automorphisms of $U(\widehat{E_n})$. Now we compute that\\

$(\nabla_{\widehat{\beta}}^0+\widehat{\Omega}_{1})(\xi)(t)$
\[~~=(\widehat{P} \otimes I_n)(\delta)(\xi)(t)+
(u{\nabla_{\beta}^0}u^*-\nabla_{\beta}^0+u{\Omega}u^*)\{\xi(t)\} \]
\[=2{\pi}it\xi(t)+\widetilde{u}\widehat{\Omega}\widetilde{u}^*(\xi)(t)+
(u{\nabla_{\beta}^0}u^*-\nabla_{\beta}^0)\{\xi(t)\} \]
\noindent
$(\xi \in \widehat{M_n(\Xi)})$ where $\widetilde{u}(\xi)(t)=u\{\xi(t)\}$. 
Since we see that\\

$(u{\nabla_{\beta}^0}u^*-\nabla_{\beta}^0)\{\xi(t)\}$
\[=(\widetilde{u}{\nabla_{\widetilde{\beta}}^0}\widetilde{u}^*-{\nabla_{\widetilde{\beta}}^0})(\xi)(t) ,\]
\noindent
and $\widetilde{u}\nabla_{\widehat{\beta}}^0\widetilde{u}^*(\xi)(t)=2{\pi}it\xi(t)$, then we obtain that 
\[\nabla_{\widehat{\beta}}^0+\nabla_{\widetilde{\beta}}^0+\widehat{\Omega}_{1})(\xi)(t)=\gamma_{\widetilde{u}}(\nabla_{\widehat{\beta}}^0+\nabla_{\widetilde{\beta}}^0+\widehat{\Omega})(\xi)(t) \]
\noindent
$(\xi \in \widehat{M_n(\Xi)})$. As we know that
\[\nabla_{\widehat{\beta}}^0+\nabla_{\widetilde{\beta}}^0
=\nabla_{\overline{\beta}}^0 \],
\noindent
then the conclusion follows. \quad Q.E.D. \\

By definition, $\widetilde{\beta}$ is a weakly inner action of $\widehat{M_n(A)}$ implimented by a unitary multiplier flow $\mu$ of $\widehat{M_n(A)}$ 
faithfully acting on $L^2(\mathbb{R},H_{\tau^n})$ for the Hilbert space 
$L^2(M_n(A),\tau^n)$. Actually, as $\widetilde{\beta}_t(a)(s)=\beta_t(a(s))$ 
for all $a \in \widehat{M_n(A)},~s,t  \in \mathbb{R}$, then $\mu_t(a)(s)=a(s-t)$ for all $a \in \mathcal{S}(\mathbb{R},M_n(A)),~s,t \in \mathbb{R}$ Then we have the following lemma: \\

Lemma 4.7 \quad The F$^*$-system $(\widehat{M_n(A)},\mathbb{R},\overline{\beta})$ is inner conjugate to the $F^*$-system $(\widehat{M_n(A)},\mathbb{R},\widehat{\beta})$. Then it implies that the $F^*$-system ($\widehat{M_n(A)}\rtimes_{\overline{\beta}} \mathbb{R},\mathbb{R},\widehat{\overline{\beta}})$ is isomorphic 
to the system $(\Doublehat{M_n(A)},\mathbb{R},\Doublehat{\beta})$ 
via the map:~$\Lambda(x)(t)=\mu_{-t}x(t)$ for all $x \in \widehat{M_n(A)}
\rtimes_{\overline{\beta}} \mathbb{R}$, where 
$\Doublehat{M_n(A)}=\widehat{M_n(A)}\rtimes_{\widehat{\beta}} \mathbb{R}$.\\

\noindent
By Lemmas 4.5 and 4.7, we deduce the following lemma: \\

Lemma 4.8  \quad Let $\Lambda_{\beta} : \mathcal{M}^{(\widehat{M_n(A)}\rtimes_{\overline{\beta}} \mathbb{R},\mathbb{R},\overline{\overline{\beta}},\Doublehat{\tau^n})}(\Doublehat{M_n(\Xi)}) \longmapsto $ \\

$\mathcal{M}^{(\Doublehat{M_n(A)},\mathbb{R},\Doublehat{\beta} \circ Ad(\nu),\Doublehat{\tau^n})}(\Doublehat{M_n(\Xi)})$ ~~defined by 
\begin{center}
$\Lambda_{\beta}([\nabla]_{U(\widehat{M_n(A)}\rtimes_{\overline{\beta}} \mathbb{R})})=[(\Lambda^n \circ \nabla \circ (\Lambda^n)^{-1}]_{U(\Doublehat{M_n(A)})}$ \end{center}
\noindent
where $\nu$ is the unitary multiplier flow of $\Doublehat{M_n(A)}$ implimenting  $\widetilde{\widehat{\beta}}$ on $\Doublehat{M_n(A)}$, and $\Lambda^n$ is the 
isomorphism from ${\mathrm{End}}_{\widehat{M_n(A)}\rtimes_{\overline{\beta}} \mathbb{R}}(\Doublehat{M_n(\Xi)})$ onto ${\mathrm{End}}_{\Doublehat{M_n(A)}}(\Doublehat{M_n(\Xi)})$ induced by $\Lambda$. Then it implies that $\Lambda_{\beta}$ 
is a homeomorphism  \\

Proof. \quad By the definition of $\Lambda$, we check that
\[ \Lambda \circ \widehat{\overline{\beta}} \circ 
\Lambda^{-1}=\Doublehat{\beta}~,~\Lambda \circ \widetilde{\overline{\beta}} 
\circ \Lambda^{-1}=\widetilde{\widehat{\beta}}.\]
\noindent
By the same reason as for $\widetilde{\beta}$, there exists a unitary multiplier flow $\nu$ of $\Doublehat{M_n(A)}$ such that $\widetilde{\widehat{\beta}}
=Ad(\nu)$ on $\Doublehat{M_n(A)}$. The rest is easily seen. \quad Q.E.D. \\

Let $\Psi_{\beta}^0$ be the isomorphism from the F$^*$-system $(\Doublehat{M_n(A)},\mathbb{R},\Doublehat{\beta})$ onto the F$^*$-system $(M_n(A) \otimes 
\mathcal{K}^{\infty}(\mathbb{R}),\mathbb{R},\beta \otimes Ad(\lambda))$ defined by
\[ \Psi_{\beta}^0(x)(t,s) = 
			\int_{\mathbb{R}} e^{2{\pi}irs}\beta_{-t}(x(t-s,r))~dr~,  \]
\noindent
and 
\[ (\Psi_{\beta}^0)^{-1}(x)(t,s) = 
		\int_{\mathbb{R}} e^{2{\pi}i(t-r)s}\beta_r(x(r,r-t))~dr   \]
\noindent
$(x \in M_n(A) \otimes \mathcal{K}^\infty(\mathbb{R}),~t,s \in \mathbb{R})$.
By definition, we compute that  \\
$\Psi_{\beta}^0 \circ Ad(\nu_p) \circ (\Psi_{\beta}^0)^{-1}(x)(t,s)$ 
\[~={\int} e^{2{\pi}i(rs+p(t-s))}\beta_{-1}((\Psi_{\beta}^0)^{-1}(x)(t-s,r))~dr~,\]
\noindent
which is equal to 
\[\int\!\!\!\int e^{2{\pi}i(p(t-s)+(t-r')r)}\beta_{r'-t}(x(r',r'-t+s))~dr'dr~.\]\noindent
$(x \in M_n(A) \otimes \mathcal{K}^\infty(\mathbb{R}),~t,s \in \mathbb{R})$. 
Therefore it follows that
\[\Psi_{\beta}^0 \circ Ad(\nu_p) \circ (\Psi_{\beta}^0)^{-1}(x)(t,s)=
   e^{2{\pi}ip(t-s)}x(t,s) ,\]
\noindent
$(x \in M_n(A) \otimes \mathcal{K}^\infty(\mathbb{R}),~p,t,s \in \mathbb{R})$,
 which implies that there exists a unitary multiplier flow $\nu_{\beta}$ of 
$\mathcal{K}^\infty(\mathbb{R})$ with the property that
\[\Psi_{\beta}^0 \circ Ad(\nu_p) \circ (\Psi_{\beta}^0)^{-1}
      =Ad(I \otimes (\nu_{\beta})_p)  \quad (p \in \mathbb{R}) \]
\noindent
on $M_n(A) \otimes \mathcal{K}^\infty(\mathbb{R})$. Then it turns out that
\[ \Psi_{\beta}^0 \circ (\Doublehat{\beta} \circ Ad(\nu)) \circ (\Psi_{\beta}^0)^{-1}=\beta \otimes Ad(\lambda \circ \nu_{\beta}) ~.\]
\noindent
Let $\Psi_{\beta}$ be the map:
$\mathcal{M}^{(\Doublehat{M_n(A)},\mathbb{R},\Doublehat{\beta} \circ Ad(\nu),\Doublehat{\tau^n})}(\Doublehat{M_n(\Xi)}) \longmapsto $ \\

$\mathcal{M}^{(M_n(A) \otimes {\mathcal{K}}^\infty(\mathbb{R}),\mathbb{R},\beta \otimes Ad(\lambda \circ \nu_{\beta}),{\tau^n} \otimes Tr)}(M_n(\Xi) \otimes {\cal{K}}^\infty(\mathbb{R})) $ \\

\noindent
induced by the equivariant isomorphism $\Psi^0_{\beta}$. Then we also show the 
following lemma by the same way as Lemma 4.8: \\

Lemma 4.9 \quad  $\Psi_{\beta}$ is a homeomorphism induced by the equivariant 
isomorphism $\Psi_{\beta}^0$. \\

Let us now consider the following map: 
\[\Pi_{\beta} : \mathcal{M}^{(M_n(A) \otimes {\mathcal{K}}^\infty(\mathbb{R}),\mathbb{R},\beta \otimes Ad(\lambda \circ \nu_{\beta}),{\tau^n} \otimes Tr)}(M_n(\Xi) \otimes \mathcal{K}^\infty(\mathbb{R})) \]
$~~~~~~~~~\longmapsto \mathcal{M}^{(M_n(A),\mathbb{R},\beta,{\tau}^n)}(M_n(\Xi))$ \\ 

defined by the natural one induced from the map $\Pi:$
\[M_n(A) \otimes \mathcal{K}^\infty(\mathbb{R}) \longmapsto M_n(A) \]
\noindent
given by $\Pi:x \mapsto (I \otimes e)x(I \otimes e)$, 
where $e$ is a rank one projection of $\mathcal{K}^\infty(\mathbb{R})$,~
$Tr$ the canonical trace of $\mathcal{K}^\infty(\mathbb{R})$.
\noindent
Now let $\nabla \in \rm{MC}(M_n(\Xi) \otimes \mathcal{K}^\infty(\mathbb{R}))$ 
and put 
\[  \nabla_e(\xi)=(I_n \otimes e)\nabla(\xi \otimes e)  \]
\noindent
$(\xi \in M_n(\Xi))$, 
where $I_n$ is the identity of $E_n$. Then $\nabla_e$ is well defined and 
independent of the choice of $e$ up to the gauge equivalence 
because of the existence of a unitary multiplier of $\mathcal{K}^\infty(\mathbb{R}))$ which sends $e$ to another rank one projection. Then we see that 
given any $u \in U(E_n \otimes \mathcal{K}^\infty(\mathbb{R}))$,
\[  (\gamma_u(\nabla))_e =\gamma_{u_e}(\nabla_e)  , \]
where $u_e = (I_n \otimes e)u(I_n \otimes e) \in U(E_n)$. Here we define a map 
$\Pi_{\beta}$ by 
\[\Pi_{\beta}([\nabla]_{U(E_n \otimes \mathcal{K}^\infty(\mathbb{R}))}
=[\nabla_e]_{U(E_n)}=[\Pi^n \circ \nabla \circ (\Pi^n)^{-1}]_{U(E_n)}   .  \]
Then it is well defined and independent of the choice of $e$. Moreover, we have the following lemma:  \\

Lemma 4.10 
\[\Pi_{\beta}:\mathcal{M}^{(M_n(A) \otimes \mathcal{K}^\infty(\mathbb{R}),\mathbb{R},\beta \otimes Ad(\lambda \circ \nu_{\beta})),{\tau^n} \otimes Tr)}(M_n(\Xi) \otimes \mathcal{K}^\infty(\mathbb{R})) \]
$~~~~~~~\longmapsto \mathcal{M}^{(M_n(A),\mathbb{R},\beta,{\tau}^n)}(M_n(\Xi))$~is a homeomorphism. \\ 

Proof. \quad Let us define the mapping $\Pi^{-1}_{\beta}$ by
\[\Pi^{-1}_{\beta}([\nabla]) ~=~ [\nabla \otimes I_{\mathcal{K}^\infty(\mathbb{R})}]   ~.   \]
Then it is easily seen that both $\Pi^{-1}_{\beta} \circ \Pi_{\beta}$ and 
$\Pi_{\beta} \circ \Pi^{-1}_{\beta}$ are identities. Moreover if 
$[\nabla^{\iota}] \longrightarrow [\nabla]$ with respect to 
$< \cdot \mid \cdot >_{M_n(A)}$, then it follows from the definition that 
there exists a unitary net $\{u_{\iota}\}$ (by choosing a subnet) of $E_n$ 
such that $\gamma_{u_{\iota}}(\nabla^{\iota}) \longrightarrow \nabla$, 
which implies that $[(\nabla^{\iota})_e] \longrightarrow [\nabla_e]$, 
so that $\Pi_{\beta}$ is continuous and so is also $\Pi^{-1}_{\beta}$ by the 
same way.   \quad  Q.E.D.  \\

Let $\nabla^0_{\cdot}$ be the Grassmann connection of $\cdot$. Then we easily 
check the following lemma: \\

Lemma 4.11
\[ \quad \Psi^0_{\beta^n} \circ \nabla^0_{\Doublehat{\beta} \circ Ad(\nu)} \circ (\Psi^0_{\beta^n})^{-1} =\nabla^0_{\beta \otimes Ad(\lambda \circ \nu_{\beta})}   ~.  \]  
Proof. \quad It follows from Lemma 4.9 that
\[  \Psi^0_{\beta} \circ (\Doublehat{\beta} \circ Ad(\nu) \circ (\Psi^0_{\beta})^{-1} = \beta \otimes Ad(\lambda \circ \nu_{\beta}) ~ . \]
\noindent
Since
\[ \Doublehat{M_n(\Xi)} = \mathcal{S}(\mathbb{R}^2,M_n(\Xi)) , \] 
and 
\[ \nabla^0_{\Doublehat{\beta}} =\Doublehat{P}~d~\Doublehat{\beta^n} ,\]
where 
\[ \Doublehat{P}=
   P \otimes I_n \otimes I_{\mathcal{S}(\mathbb{R}^2)} ,  \]
\noindent
and $d~\Doublehat{\beta^n}$ is the infinitesimal generator of 
$\Doublehat{\beta^n}$, then this implies the conclusion. \quad Q.E.D. \\

\noindent
Moreover, we need the following lemma which is directly shown: \\

Lemma 4.12  \quad There exists a 
\[U \in U(\mathrm{End}_{M_n(A) \otimes \mathcal{K}^{\infty}(\mathbb{R})}(M_n(\Xi) \otimes {\mathcal{S}}(\mathbb{R}^2))) \]
such that
\[ \Psi^0_{\beta^n} \circ \Lambda^n \circ \Doublehat{\Omega} \circ (\Psi^0_{\beta^n}\circ \Lambda^n)^{-1}=\gamma_{U}(\Omega \otimes Id) \]  
\noindent
Actually, $U$ is defined as $U(\xi)(s,t)=u^n_{-s}\xi(s,t)$ 
for all $\xi \in M_n(\Xi) \otimes {\mathcal{S}}(\mathbb{R}^2)$, 
where $u^n$ is the unitary flow to $E_n$ induced by $\beta^n$. \\

Proof. \quad We know by definition that
\[ \Psi^0_{\beta^n}(\xi)(s,t) ~= 
	 {\int}_{\mathbb{R}}~e^{2{\pi}irt}~u^n_{-s}\xi(s-t,r)~dr   \]
\noindent
$(\xi \in \Doublehat{M_n(\Xi)}$, and
\[ (\Psi^0_{\beta^n})^{-1}(\xi)(s,t) ~= 
	\int_{\mathbb{R}}~e^{2{\pi}i(s-r)t}~u^n_r\xi(r,r-s)~dr   \]
\noindent
$(\xi \in M_n(\Xi) \otimes \mathcal{S}({\mathbb{R}}^{2}))$. Then we compute that
\[ \Psi^0_{\beta^n} \circ \Lambda^n \circ \Doublehat{\Omega}(\xi)(s,t)=
\int~e^{2{\pi}irt}u^n_{-s}(\Lambda^n \circ \Doublehat{\Omega}(\xi)(s-t,r))~dr \]\noindent
$(\xi \in \Doublehat{M_n(\Xi)})$. Then as we know that
\[\Lambda^n \circ \Doublehat{\Omega}(\xi)(s-t,r)=\nu^n_{-r}\widehat{\Omega}\{\xi(r)\}(s-t)=\Omega\{\xi(s-t+r,r)\} ~.\]
Therefore we have that \\

$ \Psi^0_{\beta^n} \circ \Lambda^n \circ \Doublehat{\Omega}(\xi)(s,t) $
\[ ~=~\int~e^{2{\pi}irt}u^n_{-s}\Omega\{(\xi)(s-t+r,r)\}~dr \]
\noindent
$(\xi \in \Doublehat{M_n(\Xi)})$. Replacing $\xi$ by $(\Psi^0_{\beta^n} \circ 
\Lambda^n)^{-1}(\xi)$, we obtain that  \\

$(\Psi^0_{\beta^n} \circ \Lambda^n)^{-1}(\xi)(s-t+r,r)$ 
\[~~~=\int_{\mathbb{R}}~e^{2{\pi}i(s-t-r')r}u^n_{r'}\xi(r',r'-s+t)~dr' \]
\noindent
$(\xi \in M_n(\Xi) \otimes \mathcal{S}(\mathbb{R}^2)$. Combining the argument 
discussed above, we deduce that \\

$(\Psi^0_{\beta^n} \circ \Lambda^n) \circ {\Doublehat{\Omega}} \circ (\Psi^0_{\beta^n} \circ \Lambda^n)^{-1}(\xi)(s,t)$ 
\[=\int\!\!\!\int~e^{2{\pi}i(s-r')r}u^n_{-s}\Omega\{u^n_{r'}\xi(r',r'-s+t)\}~dr'dr ~,\]
\noindent
which is equal to 
\[u^n_{-s}\Omega\{u^n_s\xi(s,t)\}=\gamma_{U}(\Omega \otimes Id)(\xi)(s,t) ~,\]
\noindent
where $U(\xi)(s,t)=u^n_{-s}\xi(s,t)$. This implies the conclusion. 
\quad Q.E.D.  \\

We next show the following lemma which seems to be essential to prove our main 
theorem: \\

Lemma 4.13
\[ \Pi_{\beta} \circ \Psi_{\beta} \circ \Lambda_{\beta} \circ \Phi_{\overline{\beta}} \circ \Phi_{\beta} = Id   \]
\noindent
on $\mathcal{M}^{(M_n(A),\mathbb{R},\beta,\tau^n)}(M_n(\Xi))$, 
where $\Psi_{\beta}$ is the homeomorphism on the moduli space induced from the 
isomorphism $\Psi^0_{\beta}$ given in Lemma 4.9. \\

Proof. \quad Let $\nabla=\nabla^0_{\beta^n}+\Omega \in \mathrm{MC}^{(M_n(A),\mathbb{R},\beta,\tau^n)}(M_n(\Xi))$. 
Then we know by Lemmas 4.11 and 4.12 that there exist a 
\[U \in U(\mathrm{End}_{M_n(A) \otimes {\mathcal{K}}^{\infty}(\mathbb{R})}(M_n(\Xi) \otimes \mathcal{S}(\mathbb{R}^2))) \]
such that
\[(\Psi^0_{\beta^n} \circ \Lambda^n) \circ \nabla^0_{\overline{\widehat{\beta}}} \circ (\Psi^0_{\beta^n} \circ \Lambda^n)^{-1} = \nabla^0_{\beta \otimes Ad(\lambda \circ \nu_{\beta})}  ,  \]
\noindent
and 
\[(\Psi^0_{\beta^n} \circ \Lambda^n) \circ \Doublehat{\Omega} \circ (\Psi^0_{\beta^n} \circ \Lambda^n)^{-1} = \gamma_{U}(\Omega \otimes Id)  ~.  \]  
\noindent
By Lemmas 4.11 and 4.12, we obtain that
\[ \Psi_{\beta} \circ \Lambda_{\beta} \circ \Phi_{\overline{\beta}} \circ \Phi_{\beta}([\nabla]_{U(E_n)})=[(\nabla^0_{\beta^n \otimes Ad(\lambda \circ \nu_{\beta})})+\gamma_{U}(\Omega \otimes Id)]_
{U(E_n \otimes \mathcal{K}^{\infty}(\mathbb{R}))}  ~. \]
\noindent
then we see that
\[(\nabla^0_{\beta^n \otimes Ad(\lambda \circ \nu_{\beta})})_e=\nabla^0_{\beta^n } \otimes e   ~,\]
\noindent
where $e$ is a rank one projection of $\mathcal{K}^{\infty}(\mathbb{R})$. 
In fact, 
we check that  \\

$ (P \otimes I_n \otimes e)d~((\beta^n) \otimes Ad(\lambda \circ \nu_{\beta}))^n(\xi \otimes e) $
\[~=(P \otimes I_n \otimes e)d~((\beta^n)^n \otimes Id)(\xi \otimes e)+
(P \otimes I_n \otimes e)d(Id \otimes Ad(\lambda \circ \nu_{\beta})^n)
(\xi \otimes e)~ . \]
\noindent
$(\xi \in M_n(\Xi))$. Then it follows that
\[ e{Ad}(\lambda_t \circ (\nu_{\beta})_t)(e)=e  . \]
\noindent
for all $t \in \mathbb{R}$. Therefore, it follows that
\[(P \otimes I_n \otimes e)d(Id \otimes Ad(\lambda \circ \nu_{\beta})^n)(\xi \otimes e)=0 \]
\noindent
$(\xi \in M_n(\Xi))$. On the other hand, we know that
\[ \gamma_{U}(\Omega \otimes Id)_e = \gamma_{U_e}(\Omega) \otimes e ~,\]
\noindent
where $U_e$ belongs to $U(E_n)$ with $U_e \otimes e=(I \otimes e)U(I \otimes e)$. By the definition of $U$, $U_e$ commutes with $\beta^n$. Consequently, 
it follows that
\[ [\nabla^0_{\beta^n \otimes e}+\gamma_{U_e}(\Omega) \otimes e]_{U(E_n) \otimes e}=[\nabla^0_{\beta^n}+\Omega]_{U(E_n)} , \]
\noindent
which deduce the conclusion.  \quad Q.E.D.  \\

\noindent
Applying Lemma 4.13 to the system 
$(\widehat{M_n(A)},\mathbb{R},\overline{\beta})$, we obtain the following 
corollary: \\

Corollary 4.14 
\[\Pi_{\overline{\beta}} \circ \Psi_{\overline{\beta}} \circ 
\Lambda_{\overline{\beta}} \circ \Phi_{\overline{\overline{\beta}}} \circ 
\Phi_{\overline{\beta}} =Id \]
\noindent
on $\mathcal{M}^{(\widehat{M_n(A)},\mathbb{R},\overline{\beta},\widehat{\tau^n})}(\widehat{M_n(\Xi)})$, where the map 
$\Pi_{\overline{\beta}}$ is a homeomorphism from \\

$\mathcal{M}^{(\widehat{M_n(A)} \otimes \mathcal{K}^\infty(\mathbb{R}),\mathbb{R},\overline{\beta} \otimes Ad(\lambda \circ \nu_{\overline{\beta}}),\widehat{\tau^n} \otimes Tr)}(\widehat{M_n(\Xi)} \otimes \mathcal{K}^\infty(\mathbb{R}))$ \\\noindent
onto \\

${\mathcal{M}}^{(\widehat{M_n(A)},\mathbb{R},\overline{\beta},\widehat{\tau^n})}(\widehat{M_n(\Xi)}) $
induced by the mapping:
\[ \widehat{M_n(A)} \otimes \mathcal{K}^\infty(\mathbb{R}) \longmapsto 
\widehat{M_n(A)}  \]
given by $x \mapsto (I \otimes e)x(I \otimes e)~. $ \\

As we have seen in Lemma 4.8, 
\[\Lambda \circ \overline{\overline{\beta}} \circ \Lambda^{-1}=
\Doublehat{\beta} \circ Ad(\nu) \]
Then the relation between $\Phi_{\Doublehat{\beta} \circ Ad(\nu)}$ and 
$\Phi_{\beta}$ is as followsF\\

Lemma 4.15
\[ \Phi_{\Doublehat{\beta} \circ Ad(\nu)}=\Lambda_{\beta} \circ (\Pi_{\overline{\beta}} \circ \Psi_{\overline{\beta}} \circ \Lambda_{\overline{\beta}})^{-1} 
\circ \Phi_{\beta} \circ \Pi_{\beta} \circ \Psi_{\beta}   \]
\noindent
on 
\[\mathcal{M}^{(\Doublehat{M_n(A)},\mathbb{R},\Doublehat{\beta} \circ Ad(\nu),
\Doublehat{\tau^n})}(\Doublehat{M_n(\Xi)}) ~.\]

Proof. \quad By Lemmas 4.8 $\sim$ 4,13 and Corollary 4.14, we have that 
\[ (\Pi_{\overline{\beta}} \circ \Psi_{\overline{\beta}}\circ \Lambda_{\overline{\beta}})^{-1} = \Phi_{\overline{\overline{\beta}}} \circ \Phi_{\overline{\beta}}~. \]
\noindent
By the equality written just above this Lemma, we see that
\[\Lambda_{\beta} \circ \Phi_{\overline{\overline{\beta}}} \circ (\Lambda_{\beta})^{-1}=\Phi_{\Doublehat{\beta} \circ Ad(\nu)} ~.\]
\noindent
By Lemma 4.13 and Corollary 4.14, $\Phi_{\overline{\beta}}$ is bijective. 
By Lemma 4.12, we have that 
\[ \Phi_{\overline{\beta}}^{-1} = \Phi_{\beta} \circ \Pi_{\beta} \circ \Psi_{\beta} ~,  \]
\noindent
which implies the conclusion.  \quad Q.E.D.  \\

\noindent
Summing up the above argument, we obtain the following lemma: \\

Lemma 4.16 \quad $\Phi_{\beta}$ and $\Phi_{\overline{\beta}}$ are 
homeomorphisms. \\

Proof. \quad By Lemmas 4.8 $\sim$ 4.15, $\Phi_{\beta}$ and $\Phi_{\overline{\beta}}$ are bijective and bicontinuous, which completes the proof. \quad Q.E.D.\\

\noindent
We then define a map $(\Phi_{\alpha^n})^{-1}$:
\[\mathcal{M}^{(M_n(A) \rtimes_{\alpha^n} \mathbb{R}, \mathbb{R}, \overline{\alpha^n}, \widehat{\tau^n})}(\widehat{M_n(\Xi)}) \]
\[\longmapsto \mathcal{M}^{(M_n(A),\mathbb{R},\alpha^n,\tau^n)}(M_n(\Xi)) \]
\noindent
defined by 
\[\Pi_{\alpha^n} \circ \Psi_{\alpha^n} \circ (\widetilde{\iota})^{-1} \circ \Lambda_{\beta} \circ \Phi_{\overline{\beta}} \circ \iota ~,\]
\noindent
where $\iota$ is the extended map on $\mathcal{M}^{(M_n(A),\mathbb{R},\alpha^n,\tau^n)}(M_n(\Xi))$ induced by $\iota_u$ in Lemma 4.3, and so is $\widetilde{\iota}$ on $\mathcal{M}^{(\widehat{M_n(A)},\mathbb{R},\overline{\alpha^n},\widehat{tau^n})}(\widehat{M_n(\Xi)})$ induced by $\widetilde{\iota_u}$, 
because the map $\iota_u$ intertwines $\overline{\alpha^n}$ and $\overline{\beta}$, and the map $\widetilde{\iota_u}$ intertwines $\overline{\overline{\alpha^n}}$ and $\overline{\overline{\beta}}$.   \\

Lemma 4.17 \quad $(\Phi_{\alpha^n})^{-1}$ is a homeomorphism. \\

Proof. \quad By Lemma 4.3 (2), there exists a unitary multiplier $W$ of 
$M_n(A) \otimes \mathcal{K}^\infty(\mathbb{R})$ such that 
\[Ad(W) \circ \Psi_{\beta}^0 \circ \widetilde{\iota_u} = \Psi_{\alpha^n}^0~,\]
\noindent
which implies that $\Psi_{\beta} \circ \widetilde{\iota} = \Psi_{\alpha^n}$. 
Moreover, $\iota$ and $\widetilde{\iota}$ are homeomorphisms. Then it follows 
from Lemma 4.16 that $(\Phi_{\alpha^n})^{-1}$ is a homeomorphism. \quad Q.E.D.\\\noindent
By Lemmas 4.17, we deduce the following corollary: \\

Corollary 4.18 \quad $\Phi_{\alpha^n}$ is a homeomorphism:
\[\mathcal{M}^{(M_n(A),\mathbb{R},\alpha^n,\tau^n)}(M_n(\Xi)) \longmapsto 
\mathcal{M}^{(M_n(A) \rtimes_{\alpha^n} \mathbb{R}, \mathbb{R}, \overline{\alpha^n}, \widehat{\tau^n})}(\widehat{M_n(\Xi)})~.\]

\noindent
Finally, using $\Phi_{\alpha^n}$, we define the map $\Phi_{\alpha}$ : 
\[\mathcal{M}^{(A,\mathbb{R},\alpha,\tau)}(\Xi) \longmapsto \mathcal{M}^{(\widehat{A},\mathbb{R},\overline{\alpha},\widehat{\tau})}(\widehat{\Xi}) \]
\noindent
by
\[\Phi_{\alpha}=\widehat{\Pi_n} \circ \Phi_{\alpha^n} \circ (\Pi_n)^{-1} ~,\]
\noindent 
where $\Pi_n$ is a homeomorphism:
\[\mathcal{M}^{(A \otimes M_n(\mathbb{C}),\mathbb{R},\alpha \otimes Ad(\lambda \circ \nu_{\alpha}),\tau \otimes Tr_n)}(\Xi \otimes M_n(\mathbb{C})) \longmapsto \mathcal{M}^{(A,\mathbb{R},\alpha,\tau)}(\Xi) ~.\]
\noindent
Finally, we show the following main lemma: \\

Lemma 4.19  \quad $\Phi_{\alpha}$ is a homeomorphism. \\

Proof.  In Lemma 4.10, replacing $(M_n(A) \otimes \mathcal{K}^\infty(\mathbb{R}),\beta \otimes Ad(\lambda \circ \nu_{\beta})$ and $M_n(\Xi) \otimes \mathcal{K}^{\infty}(\mathbb{R})$ by $A \otimes M_n(\mathbb(C)),\alpha^n$ and $\Xi \otimes 
M_n(\mathbb{C})$ respectively, we deduce that both $\widehat{\Pi_n}$ and 
$(\Pi_n)^{-1}$ are homeomorphisms. Then it implies the conclusion 
by Corollary 4.18. \quad Q.E.D.  \\

\noindent
Summing up all the argument discussed above, we obtain the main result of 
Theorem 4.1  \\

In what follows, we compute the moduli spaces of some concrete examples 
by means of Theorem 4.1 : \\

Example 4.20  
\[\mathcal{M}^{(\mathcal{K}^{\infty}(\mathbb{R}),\mathbb{R},Ad(\lambda),Tr)}
({\mathcal{K}}^{\infty}(\mathbb{R})) \approx {\mathcal{M}}^{({\mathcal{S}}(\mathbb{R}),\mathbb{R},\lambda,\int)}(\mathcal{S}(\mathbb{R}))  \]
\[ \qquad~~~~~~~~~~~~~\qquad ~~~~~\quad ~~\approx {\mathcal{M}}^{(\mathbb{C},\mathbb{R},Id,1)}(\mathbb{C}) \approx \mathbb{R} , \]
where ~$\approx$~ means a symbol of homeomorphicity.  \\

Examples 4.21 \quad Given a $\theta \in \mathbb{R}$, let us take the Moyal 
product $\star_{\theta}$ on $\mathcal{S}(\mathbb{R}^2)$. Then $(\mathcal{S}(\mathbb{R}^2),\star_{\theta})$ becomes a $F^*$-algebra, which is denoted by $\mathbb{R}_{\theta}^2$. Since $\mathbb{R}_{\theta}^2$ is isomorphic to $\mathcal{S}(\mathbb{R}) \rtimes_{\theta} \mathbb{R}$, then it follows from Theorem 4.1 that
\[{\cal{M}}^{(\mathbb{R}_{\theta}^2,\mathbb{R},\overline{\theta},\tau_{\theta})}(\mathbb{R}_{\theta}^2) \approx \mathbb{R}  , \]
\noindent
where $\tau_{\theta}$ is the canonical trace of $\mathbb{R}_{\theta}^2$. \\

\noindent
Even though changing F$^*$-flows into F$^*$-multiflows, the same result as 
Theorem 4.1 is obtained by using the ideas developed in Lemmas 4.3 $\sim$ 4.18 
as follows:  \\

Theorem 4.22  \quad Let $(A,\mathbb{R}^n,\alpha)$ be a F$^*$-multiflow with a 
faithful continuous $\alpha$-invariant trace $\tau (n \geq 1)$, and $\Xi$ a 
finitely generated projective $A$-module. Then there exist a F$^*$-multiflow 
$(\widehat{A},\mathbb{R}^n,\overline{\alpha})$ with a dual trace 
$\widehat{\tau}$, and a dual $\widehat{A}$-module $\widehat{\Xi}$ such that
\[{\mathcal{M}}^{(A,\mathbb{R}^n,\alpha,\tau)}(\Xi) \approx {\mathcal{M}}^{(\widehat{A},\mathbb{R}^n,\overline{\alpha},\widehat{\tau})}(\widehat{\Xi})~. \]

Proof. \quad Suppose $n=2$, we may choose a cocycle unitary multiplier with the property in Lemma 4.3 because $\alpha(\mathbb{R}^2)$ is a commutative connected Lie group. We then choose an outer equivalent F$^*$-multiflow $(M_n(A),\mathbb{R}^2,\beta)$ of $(A,\mathbb{R}^n,\alpha)$ such that $\beta_{(t,s)}(P)=P$ for all $(t,s) \in \mathbb{R}^2$. Then we see that the Grassmann connection $\nabla^0_{\beta}$ satisfies the Yang-Mills condition. In fact, since $\beta_{(t,s)}(P)=P$ 
for all $(t,s) \in \mathbb{R}^2$, it follows that \\
$ ~~~~~~~\Theta_{\nabla^0}(X,Y) = \nabla^0_X\nabla^0_Y-\nabla^0_Y\nabla^0_X $ 
\[~~~~~~~~~~~~~~~~~~~~=~P^n(d\beta^n)_XP^n(d\beta^n)_Y-P^n(d\beta^n)_YP^n(d\beta^n)_X  \]
$(X,Y \in \mathbb{R}^2)$, where $P^n = P \otimes I_n$. As $d\beta^n$ is a Lie 
homomorphism and $(d\beta^n)_X(P^n)=0 ~(X \in \mathbb{R}^2)$, we have that 
\[{\Theta}_{\nabla^0}(X,Y) = P^n\{(d\beta^n)_X(d\beta^n)_Y-(d\beta^n)_Y(d\beta^n)_X\} = 0  \]
$(X,Y \in \mathbb{R}^2)$. Then it follows from Theorem 2.1 that $\nabla^0$ is a  Yang-Mills connection. By the same method as in the proof of Theorem 4.1, we 
deduce the conclusion. The way used above is also applicable to the case for $n\geq 3$ by induction.  \quad Q.E.D.  \\
\noindent
The similar statement to Corollary 4.2 is in the following: \\

Corollary 4.23 \quad Let $(A,\mathbb{R}^n,\alpha)$ be a F$^*$-multiflow with a 
faithful continuous $\alpha$-invariant trace $\tau$, and $(\widehat{A},\mathbb{R}^n,\overline{\alpha})$ its associated F$^*$-flow with the dual trace $\widehat{\tau}$. Suppose $(\widehat{A},\mathbb{R}^n,\beta)$ is another $F^*$-multiflow such that 
\[\widehat{\tau} \cdot \beta = \widehat{\tau} , ~\beta \circ \overline{\alpha} = \overline{\alpha} \circ \beta ~\]
\noindent
then given a finitely generated projective $\widehat{A}$-module $\Xi$, there exist a F$^*$-multiflow $(A,\mathbb{R}^n,\beta_{A})$, a finitely generated projective $A$-module $\Xi_{A}$ and a faithful $\beta_{A}$-invariant trace $\tau_{A}$ of $A$ such that 
\[\mathcal{M}^{(\widehat{A},\mathbb{R}^n,\beta,\widehat{\tau})}(\Xi) \approx 
 \mathcal{M}^{(A,\mathbb{R}^n,\beta_{A},\tau_{A})}(\Xi_{A})  .  \]

\pagebreak

\begin{center}
Acknowledgement
\end{center}

I would like to express my sincere gratitude to Professor T.Natsume for his 
careful reading and many pieces of advice to my manuscript, and to Ms.J.Takai 
for her constant encouragement.

\vspace{1cm}

\begin{center}
References
\end{center}

\vspace{1cm}
\noindent
[1]~A.Connes, An Analogue of the Thom Isomorphism for Crossed 
Products of a C*-Algebra by an Action of $\mathbb{R}$, Adv.Math.,39,(1981),
\linebreak
31-55. \\
\noindent				  
[2]~A.Connes, M.R.Douglas and A.Schwarz, Noncommutative Geometry and Matrix 
Theory: Compactification on Tori, JHEP, 9802(1998)003.\\
\noindent				  
[3]~A.Connes and M.A.Rieffel, Yang-Mills for noncommutative two tori,
Contemp.Math.Oper.Alg.Math.Phys.62,AMS~(1987), 237-266.\\
\noindent
[4]~~Y.Ohkawa, ~Matrix Models of M-Theory, 
Math.Sci.,4(2002),
\linebreak
35-40. \\
\noindent
[5]~N.Ohta, Duality of Superstring Theory and M-Theory, 
Math.
\linebreak
Sci.,4(2002),16-22. \\
\noindent
[6]~H.Kawai, Constructive Formulation of String Theory, 
Math.Sci.,
\linebreak
4(2002),41-48. \\
\noindent
[7]~G.A.Elliott,T.Natsume and R.Nest, Cyclic cohomology for one parameter 
smooth crossed products, 
Acta Math.,160(1988), 285-305. \\

\end{document}